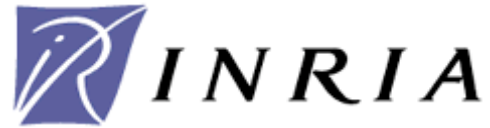

INSTITUT NATIONAL DE RECHERCHE EN INFORMATIQUE ET EN AUTOMATIQUE

# Artificial table testing dynamically adaptive systems

Freddy Munoz, Benoit Baudry

N° 6866

March 2009

Thème COM

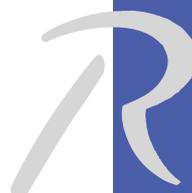

# Artificial table testing dynamically adaptive systems

Freddy Munoz[1], Benoit Baudry[2]

**Abstract:** Dynamically Adaptive Systems (DAS) are systems that modify their behavior and structure in response to changes in their surrounding environment. Critical mission systems increasingly incorporate adaptation and response to the environment; examples include disaster relief and space exploration systems. These systems can be decomposed in two parts: the adaptation policy that specifies how the system must react according to the environmental changes and the set of possible variants to reconfigure the system. A major challenge for testing these systems is the combinatorial explosions of variants and environment conditions to which the system must react. In this paper we focus on testing the adaption policy and propose a strategy for the selection of environmental variations that can reveal faults in the policy. Artificial Shaking Table Testing (ASTT) is a strategy inspired by shaking table testing (STT), a technique widely used in civil engineering to evaluate building's structural resistance to seismic events. ASTT makes use of artificial earthquakes that simulate violent changes in the environmental conditions and stresses the system adaptation capability. We model the generation of artificial earthquakes as a search problem in which the goal is to optimize different types of environmental variations.

[1] INRIA – freddy.munoz@inria.fr
[2] INRIA – benoit.baudry@inria.fr

# Artificial table testing dynamically adaptive systems

**Résumé:** Dynamically Adaptive Systems (DAS) are systems that modify their behavior and structure in response to changes in their surrounding environment. Critical mission systems increasingly incorporate adaptation and response to the environment; examples include disaster relief and space exploration systems. These systems can be decomposed in two parts: the adaptation policy that specifies how the system must react according to the environmental changes and the set of possible variants to reconfigure the system. A major challenge for testing these systems is the combinatorial explosions of variants and environment conditions to which the system must react. In this paper we focus on testing the adaption policy and propose a strategy for the selection of environmental variations that can reveal faults in the policy. Artificial Shaking Table Testing (ASTT) is a strategy inspired by shaking table testing (STT), a technique widely used in civil engineering to evaluate building's structural resistance to seismic events. ASTT makes use of artificial earthquakes that simulate violent changes in the environmental conditions and stresses the system adaptation capability. We model the generation of artificial earthquakes as a search problem in which the goal is to optimize different types of environmental variations.

**Mots clés:** Testing of dynamically adaptive systems, Artificial shaking table testing, Adaptation policy testing, Search based testing.

## 1 Introduction

Software is expected to do more for us today in more situations than we ever expected in the past. Nowadays, there exist more users, interacting systems, resources and goals than before. That is translated into system operating non-stop on complex, rapidly changing, and possibly hostile environments. It is unacceptable for these systems to crash when confronted with changes; they must instead fluidly adapt to the ongoing circumstances and find the way to continue accomplishing their functionalities. Such systems, called *dynamically adaptive systems (DAS)*, play increasingly vital roles in society's infrastructures. The demand for *DAS* appears in application domains ranging from crisis management applications such as disaster management [17], space exploration [12], and transportation control to entertainment and business applications such as mobile interactive gaming and business collaborations (e.g., through virtual organizations and dynamic service compositions). This demand is accentuated by the mobile and nomadic nature of many of these domains. Indeed, future applications will need to cope with advanced properties such as context awareness and mobility. The IDC[3] analysts forecast a global increase in the number of mobile workers to more than 850 million by 2009 [10].

*DAS* responds to environmental changes by modifying their internal configuration in order to continue meeting their functional and non-functional requirements.

Designing a *DAS* consists in two phases. The first is the identification of the system parts that may vary during the execution. Typically, designers address this step using software product lines techniques (*SPL*) [3]. *SPL* proposes to define a family of software starting from a core system and encoding its varying parts into variation points. These variation points enable determine the system's changeable structure while maintaining its overall organization. In this way, the structural changes performed in adaptation are reflected as the transition between variants of the system. During the second step, designers specify which environmental fluctuations should have an impact on the system as well the associated strategies to perform the structural changes. Typically, designers address this step by defining *adaptation policies* that encode the courses of actions to be adopted when the environment changes [2, 13, 18, 21]. Adaptation policies drive the adaptation process and compute the right system variant that should be adopted given an environmental condition.

We distinguish two activities for testing an adaptive system. The first activity consists in testing the system variants. That is, for each variant, a set of test scenarios is executed to check the variant validity. However, due to the exponential growth of system variants with the number of variables, it is impossible to perform this activity for all of them. Instead, it is necessary to select a representa-tive subset of variants to be tested. Existing testing techniques for software product lines can be applied to address this issue [5].

---

[3] IDC is an analyst company and a global provider of market intelligence, advisory services, and events for the information technology, telecommunications, and consumer technology markets

The second activity consists in testing whether the adaptation policies are correctly implemented and well suited for their working environment. That is, simulate environmental changes and check whether the system adapts correctly with respect to those changes and with respect to the adaptation policy. Doing so is challenging because again there is a problem lying on exponential growth. Simulating environmental changes requires moving the environment from one condition to another (environmental transition). Simulating the whole environment is impossible due to: (i) the extremely large number environmental conditions, and (ii) the even larger number of environmental transitions.

In this paper we address the selection of representative environmental conditions, and environmental transitions to test the adaptation policies of *DAS*. Our strategy is based on the metaphor of a civil engineering testing technique, where structural engineers test the structural resistance of building by simulating natural *earthquakes*. This kind of test is referred as *shaking table testing (STT) [11]*, because it involves placing a structure scale model over a table capable of oscillating in such a frequency and cadence that simulate a natural earthquake. Analogous to *STT* we propose *artificial shaking table testing (ASTT)* for testing adaptation policies and their realization. *ASTT* consists in laying a *DAS* into a *virtual shaking table*, which produces *artificial earthquakes (AEQ)* that test its adaptation capabilities. *AEQs* are series of environmental conditions, where at least two consecutive conditions are very different, i.e. series with strong and smooth environmental variations.

Generating *AEQs* embodies several challenges: (a) selecting series of environmental conditions that are consistent with the real occurrence of the environment; (b) selecting as much series as necessary to cover a testing criterion for adaptive systems; (c) selecting series containing as much violent variations as possible.

In order to address these challenges, we model the generation of *AEQs* as a functional optimization problem that consists in optimizing the compromise between the previous challenges. This allows us to adapt existing search-based techniques such as hill-climbing, tabu-search and simulated annealing to automatically generate *AEQs*. The virtual table is now a set of optimization goals and search algorithms.

The contribution of this paper is a technique to automatically generate *AEQs* in such a way that they simulate representative environmental changes. The experimental results of performing mutation analysis over an adaptive web server indicate that automatically generating violent and smooth environmental variations are beneficial to uncover faults in adaptation policies and their realization.

The reminder of this paper proceeds as follows. Section 2 gives a background on dynamically adaptive systems. Section 3 describes the challenges in testing adaptation policies. Section 4 introduces *artificial shaking table testing*. Section 4 presents the results of an experimental study over an adaptive web server.

Section 5 presents the related work. Finally, in section 6 we conclude and present our perspectives.

## 2 Dynamically adaptive systems

Consider a simple adaptive web server, which processes file requests over the http protocol. It answers the requests it receives as fast as possible while optimizing the resources it consumes. Additionally, it provides a non-stop service and thus it needs to modify its internal structure in order to respond to its changing working environment. The working environment of the web server is characterized by the variable amount of requests over time.

### 2.1 Environment and variants

Dynamically adaptive systems (*DAS*) encode the environment into an abstraction called a *context*.

DEFINITION 2.1. *A context consists on an n-tuple of fields $<p_1, p_2, ..., p_n>$, where each field $p_i$ represents an environmental property. The type of each field is defined by the encoding chosen for the property it represents.*

In our adaptive web server example, the environment is modeled as a context with the properties $p_1$: *number of request per second (request density)*; $p_2$: *the amount of files that can be requested (file number)*; and, $p_3$: *dispersion of the request (request dispersion)*. The last one corresponds to the percentage of requests that point to different files (among *file number*). The domain or *type* of each property has a lower and an upper bound. For instance, *Request density* and *file number* are integer numbers with lower bound 1 and upper bound 1000, whereas *request dispersion* is a real number with period 0.1, lower bound 0 and upper bound 1. The *request density* domain indicates that the minimum amount of request in one second is 1 and the maximal is 1000. Analogous, *request dispersion* indicates that when every request points the same file it has a value 0, and when all the requests are uniformly distributed among the possible files has a value 1.

DEFINITION 2.2. *Specific environmental conditions at an instant t are drawn by an instance I of the context representing the environment. Such instance is an n-tuple of values corresponding to the punctual value of a particular property.*

The context instance *<12, 3, 0.5>* designates a particular environmental condition, where 12 files are requested each second, the requests point to 3 different files, and out of 12 requests 6 point the same file. A series of context instances $I_0, I_1, I_2, ..., I_n$ ordered by their occurrence over time is called *context flow (F)*.

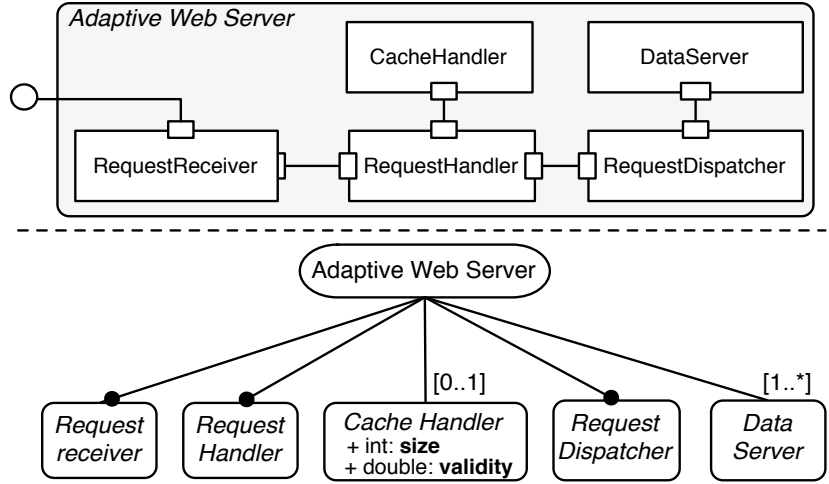

**Figure 1: Architectural and feature diagram view of the simple adaptive web server.**

Since the adaptive web server works on a changing environment, it dynamically modifies its internal structure to continue running. Figure 1 presents different views of the web server structure. On the top, an architectural view describes the structure in terms of components and connections. It states that the web server is composed of a single *request receiver*, which receives the http requests, encodes and passes them to a *request handler*. The request handler verifies through a *cache handler* whether the requested file exists in an optional cache. If the document exists, then the request is immediately answered. Otherwise the handler passes the request to one of the available *data severs*, which loads the file from a resource and answer the request. On bottom, a feature diagram [9] describes the software product line (*SPL*) comprising the different variation points of the web server. More precisely, it states that a web server must deploy exactly one (black dot) *request receiver*, one *request handler*, one *request dispatcher*, optionally one *cache handler* (0 or 1), and any number of *data servers* (1 or more). Besides, when a cache handler is deployed, the *size* and duration *validity* (seconds the files are present) of the cache must be specified.

The variation in *DAS* is commonly represented by an abstraction that encodes the system variants.

DEFINITION 2.3 *Variation is a n-tuple of field $<v_0, v_1, ..., v_n>$, where each field $v_i$ corresponds to a variation point (actually varying). Analogously to environmental properties, each field bears a type encoding the variation point they represent.*

The variation of the adaptive web server includes the variation points previously described: $v_1$: *cache existence*, $v_2$: *cache size*, $v_3$: *cache duration validity*, and $v_4$: *amount of data servers*. The domain for *cache existence* is a Boolean indicating that either the cache exists or not. When the case exists, its size *cache* *size* varies between 10 and 1024. The number of data servers varies between 1 and 100, which means that always must be at least one data server and no more than 100.

DEFINITION 2.4. *A specific configuration of the system at a moment t is drawn by a system variant δ. Each value in δ matches to the variation points values selected for that particular variant.*

The system variant *<true, 10, 2,1>* designates a configuration with a cache of size 10, a duration of 2 seconds per file, and only one data server. Analogously to a context flow, a *variant flow* reflects the configuration changes over time.

Context and variation raise a space containing all the possible instances / variants that can produce the combination of the properties / variation point values. For example, the context of the adaptive web server raises a space containing all its possible context instances.

DEFINITION 2.5 *Context instances (I) as well as system variants (δ) must satisfy a series of constraint (ς) specific to their encoding and domain. Any context instance or system variant violating these constraints is invalid and does not belongs to the context / variant space they represent.*

Two constraints are defined for the adaptive web server. The first is a constraint on the variant space and states that *when a cache exists its size and duration must be great than 0, otherwise they should be 0*. An example of a system variant violating this constraint is *<false, 10, 2,1>*. The second constraint restricts the context space and states that *the number of files that can be requested cannot be superior to the number of requests per second*. That is, the number of possible files grows linearly with the number of requests per second.

## 3 Adaptation driver

Adaptation in *DAS* is driven by a series of adaptation policies (adaptation model) that use different formalisms to describe the variant to adopt given a context change.

DEFINITION 2.6. *An adaptation policies fp defines a relation between context and system variants. It is a function fp: ℘ (I) x ℘ (δ) → δ that receives a context flow (context history), a variant flow (variation history), and gives the next variant the system must adopt.*

There exist several strategies to implement adaptation policies, a few examples are: action-based adaptation [18], where adaptations are triggered when a condition is satisfied; goal based adaptation [13], where adaptations are performed to reach a specific goal; and utility function based adaptation [21], where adaptations are calculated according to a cost function based on environmental conditions and variation point value.

The adaptation policies of the adaptive web server use a class of action-based strategy [18]. In this case the adaptation policy is a set of rules that, for each event (environmental change) evaluate if a set of conditions are satisfied, and if it is so, they perform a series of adaptation actions. In particular, this strategy encodes the condition values using fuzzy logic transformations [23]. This enable designers to write conditions based on adjectives (fuzzy values) such as *high, medium, low* instead of precise value. The use of such adjectives allows designers to abstract from low-level details and qualitatively define the system adaptations [2].

**Listing 1: Excerpt of the adaptive web server adaptation rules.**

Listing 1 presents an excerpt of the adaptive web server adaptation policy[4]. It contains 2 rules (lines 1-3 and 5-7), which state the utility of adding a cache (adaptation action), given certain request dispersion. The first (lines 1-3) states that whenever the request dispersion is *low* or *medium* (line 1) and there is no cache (line 2), deploying a cache is *very useful* (line 3, value *high*). The utility of adding a cache is also an adjective, in this way the adaptation policy remains abstract from the application domain. Later on, a fuzzy engine assigns numerical values to this adjective and whether it reaches a threshold, the system deploys a cache. The second rule (lines 5-6) is analogous to the first; nonetheless, it states that when the dispersion is high, adding a cache is not very useful.

## 4 Testing the adaptation policy

Definition 2.6 introduces the concept of adaptation policy as the driver of the adaptation. Testing the realization of such driver means verifying whether the system is capable of adapting to environmental changes, and whether such adaptations proceed as specified in the adaptation policy. Additionally, tests assess the adequacy of adaptation policies with respect to the possible environmental changes. That is, they can help uncovering unforeseen environmental conditions that may not be covered by the adaptation policy.

Testing adaptation policies involves generating context instances, and evaluating the results of exposing the system to such context instances.

Figure 2, illustrates the testing process for adaptation policies. It is composed of the three steps. (1) Initially, testers synthesize a context flow from a series of

```
1: WHEN REQUESTDISPERSION IS 'LOW' OR 'MEDIUM'
2: IF CACHEHANDLER.ISEMPTY
3: THEN UTILITY OF ADDCACHE IS 'HIGH'

5: WHEN REQUESTDISPERSION IS 'HIGH'
6: IF CACHEHANDLER.ISEMPTY
7: THEN UTILITY OF ADDCACHE IS 'LOW'
```

---

[4] The full adaptation policy can be found at http://freddy.cellcore.org/research/cherokee/rules.html

context instances. (2) Then, they execute and expose the system to the generated context flow. That is, varying the system environment as described by the instances in the flow. Additionally, tester must analyze the adaptation policy in order to calculate the expected variant for each instance in the context flow; this generates an expected variant flow. (3) Finally, testers evaluate whether the variants adopted by the system (variant flow) when exposed to environmental changes are equivalent to the expected variant flow. If it is the case the process may start again until a stop criterion is reached. Such criterion may be for instance the coverage of the whole environment. Otherwise, there is a fault and must be localized and corrected.

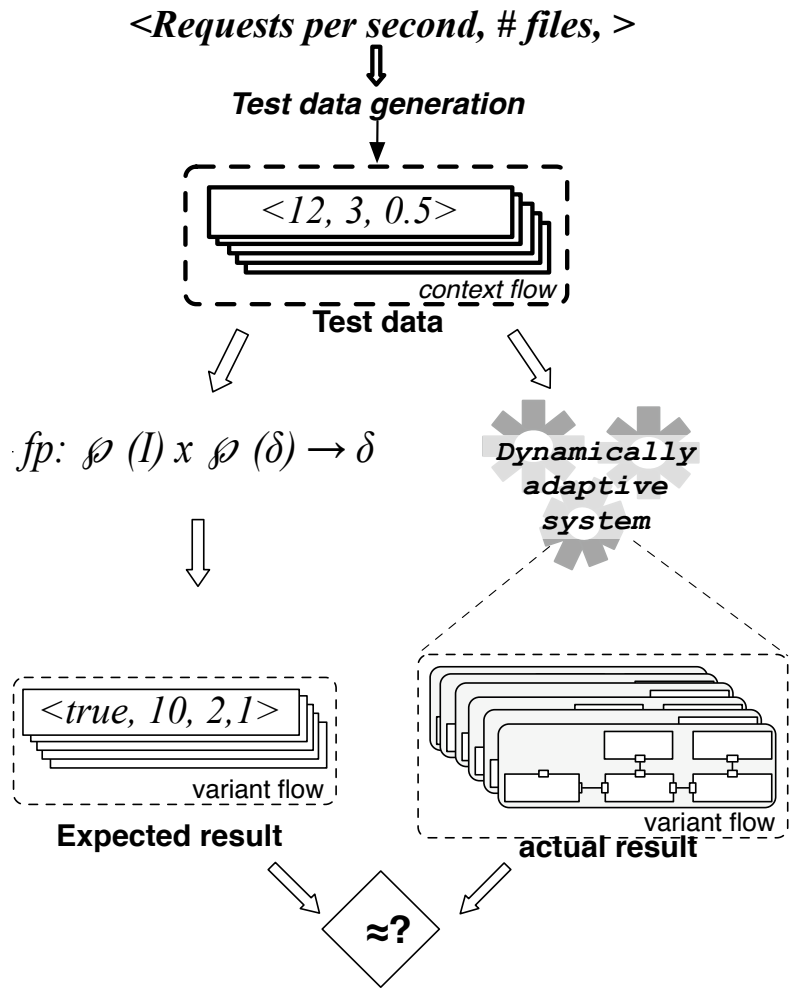

Figure 2: Testing of adaptation policies

Executing the testing process previously described raises two challenges. The first challenge relates to generating test data. Ideally we may want to generate all the context instances in order to assure the validity of adaptation policies over the whole environment. However, the number of context instances (context space) grows exponentially with the number of properties modeled by the context. For example, the context of the adaptive web server models 3 properties that generate a space of $10^6$ (1000 x 10 x 500) instances. Moreover, since adaptation policies use the history of the context flow that has occurred before the new context instance, the configuration chosen by the adaptation policy may change according to the system history. It is possible that context flows containing identical context instances, but with different order of occurrence over time may lead to completely different variants. For this reason, the order of occurrence of context instances in a context flow cannot be disregarded. This creates a *flow space* containing the different context flows of a context. The length *n* of the flows determines the size of such space, which is equal to *j x k*, where there are *k* possible ways to choose *n* values from the context space, and *j* possible combinations of *n* context instances *(j=n x (n-1))*. For example, the size of the adaptive web server's flow space of the, with context flows of length $10^6$ is $10^{12}$ ($10^6$ x ($10^6$-1) x 1). This means that for testing the adaptation policies of our adaptive web server against all the possible environmental variations it is necessary to execute the testing process at least $10^{12}$ times. Rapidly we notice that executing such amount of tests is not feasible in reasonable time. Synthesizing context flows is challenging because it involves selecting the minimal representative amount of context instances and arranging them in the right way to produce context flows that adequately represent environmental variations.

The second concerns the evaluation of the variant flow produced by the system. Generally adaptation policies are hard-coded into the system without a proper specification. Since there is no proper specification, it is difficult to calculate the expected variant flow. Moreover, some adaptation policies may not produce deterministic results, and rather produce a set of possible results or template results. In such cases, evaluating whether the expected and the produced variant flow are equivalent is complex because they may share only some commonalities.

In this paper we propose a strategy to explore the context flow space and to cope with the first challenge. Inspired by civil engineering structural testing we propose to synthesize context flows with particular properties that stress the system's environment. Such flows may exercise the adaptation policy and uncover faults. Concerning the evaluation of the expected result, we propose a solution specific to our case study.

## 5 Artificial shaking table testing

Consider the following scenario for the adaptive web server. Initially, the server receives 10 requests per second equally distributed between 5 files, has a configuration of 1 file server and no cache deployed. Suddenly, it receives 1000 requests per second distributed between 300 files of which 2 repeats 300 times. The adaptation policy specifies that the server should deploy as fast as possible the cache with a sufficient size to hold the 300 files, and enough file servers to fetch 300 files. After a few seconds, the server request rate returns to its initial value and the server should remove the deployed cache and file servers in order to save resources.

This scenario is an example of violent changes in the environmental conditions of a *DAS*. The realization of the adaptation policy should drive the system's adaptation responding to such changes and produce the described configuration changes. However, if it fails to do so, the system may not meet its functional and nonfunctional requirements, and will not be able to provide the expected service. We can speculate about a variety of faults that lie on the adaptation policy realization or specification, and that can be revealed by the described scenario. Three faults that may occur are the following. (1) Consider that the adaptive web server is faced to the described scenario, but, when required, it does not deploy the file servers, or the cache, or the size of the cache is too small to hold 2 different files. If request response quality of service requirements were bounded to the system, it will not be able to meet them. Since the adaptation policy specification stipulates the way the adaptation must proceed, the source of misbehavior is located in the policy realization. (2) Consider that the adaptive web server is capable of deploying the file servers and the cache with the right size, but with a long delay. Again, if quality of service requirements were bounded to the system, it will not be able to meet them because the response time will be too long, and probably when deployed, the file server and the cache will be useless. (3) Now, consider that the adaptive web server actually deploys the servers and the cache as needed, but it does not remove them after the request rate descends. If memory and calculation resources are scarce, the server will be over-consuming them and eventually crash. This misbehavior can be due to a faulty realization of the adaptation policy, or to a fault in the adaptation policy; for example, the adaptation policy does not specify what to do when the server request rate decreases, or there exist some contradictory rules.

In the previous section we stated the challenges of testing the adaptation policy realization. Particularly we highlighted the challenge of synthesizing flows representative of the context flow space. Such context flow must aim at uncovering faults due to violent environmental changes and ensure that the system is capable of adapting in violent and non-violent environment.

We address this issue by proposing a strategy to synthesize context flows, which contain violent and non-violent environmental changes. Based on the metaphor of a civil engineering testing technique referred as *shaking table testing*, we propose to generate *artificial earthquakes*: context flows with several violent environmental changes. Our hypothesis is that synthesizing context flows with the described property will help testers finding faults in the adaptation policy implementation and design. Additionally, such context flows may help developers checking whether the DAS meets its requirements. For instance, whether the adaptation is carried within 5 seconds.

In the reminder of this section we present the underlying idea of shaking table test to later introduce our strategy to artificial shaking table testing.

### 5.1 Shaking table testing & Artificial shaking table testing

*Shake table testing* (*STT*), or earthquake testing is a technique widely used in civil engineering to test the structural resistance of buildings to ground movements such as earthquakes. It consists in simulating the shaking effect of earthquake over a target structure. It uses a table (*shaking table*) that sustains the structure under test and oscillates with different intensities and cadence rates over time; this produces waves that stress the tested structure's material resistance and design [11]. *STT* is used to test the structural integrity, construction material, and structural configuration of a building facing the effects of an earthquake or another ground movement. *STT* helps civil engineers to develop structures that better resist natural disasters such as earthquakes without risking human lives in the process.

Analogous to *STT*, we propose using a *virtual* shaking table to test the *resistance* of adaptation policies to smooth and violent environmental changes. We refer to such testing strategy as *artificial shaking table testing (ASTT)*. ASTT operates by generating data we refer as *artificial earthquakes*, in reference to natural earthquakes that produce sudden and violent ground movement. Artificial earthquakes are context flows embodying violent and sudden context changes, which are transitions between two instances located as far as possible from each other in the context space. Our hypothesis is that such changes may stress the implementation of the adaptation policy and help testers uncovering faults related to transitions between context instances with different degree of separation. Furthermore, through the exploration of the context flow space, artificial earthquakes may help testers uncovering design faults in the adaptation policy specification and therefore assess their adequacy with respect to their working environment.

### 6 Artificial Earthquakes

*Artificial shaking table testing (ASTT)* uses artificial earthquakes as the core element for testing an adaptation policy. We define an artificial earthquake and its component elements as follows.

DEFINITION 5.1 *An artificial earthquake (AEQ) Æ is a context flow f, which exhibits an earthquake profile (EP).*

An *earthquake profile EP* is the fundamental property of an *AEQ*, and, as its name suggests, it is the presence of a *virtual earthquake* among the elements of a context flow. Consider a context flow *f* composed of a series of context instances $I_0, I_1, ..., I_n$, ordered in such a way that at some point the distance between a series of consecutive instances increases violently in relation with the prior distances. We call such violent variation of the distance between consecutive context instances and *EP*. A precise definition of an EP relies on the definition of distance between a pair of context instances.

DEFINITION 5.2 *The distance between two context instances $I_i$, $I_{i+1}$ is defined as $D(I_i, I_{i+1})$ where $D : I \times I \rightarrow \mathbb{R}$ is a function that assigns a distance (continuous value) to a pair of context instances.*

The function $f_d$ depends on the application domain of the system under test. It basically maps a pair of context instances (tuples of values) into a single value representative of their location in the context space. For instance, such function for the adaptive web server context corresponds to the Euclidian distance between two triplets of values.

DEFINITION 5.3 *The origin $\otimes$ of a context space is a single instance that represents a reference point in the context space.*

The origin of the context space gives us a stand ground to define *EP*, and allows us to state important properties that an *EP* must have.

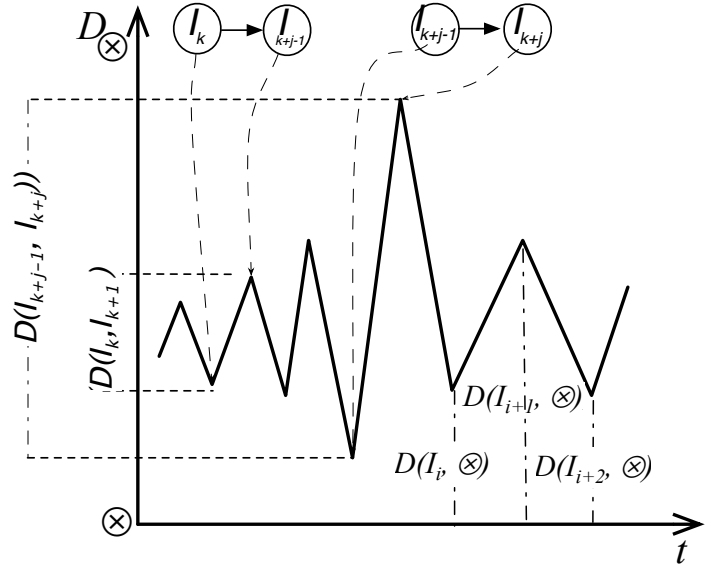

**Figure 3: Graphical representation of a context flow with an *EP***

DEFINITION 5.4 *An earthquake profile in a context flow **f** consists in a violent variation of the distance between two or more context instances.*

(1) $\exists <I_k, I_{k+1},...,I_{k+j}> \subseteq f$ /
$(D(I_k, I_{k+1}) << D(I_{k+j-1}, I_{k+j})) \vee$
$(D(I_k, I_{k+1}) >> D(I_{k+j-1}, I_{k+j}))$, $k=1..n$

(2) $\forall I_i, I_{i+1}, I_{i+2} \in f$,
$(D(I_i, \otimes) \geq D(I_{i+1}, \otimes) \wedge D(I_{i+1}, \otimes) \leq D(I_{i+2}, \otimes)) \vee$
$(D(I_i, \otimes) \leq D(I_{i+1}, \otimes) \wedge D(I_{i+1}, \otimes) \geq D(I_{i+2}, \otimes))$, $i=1..n$

*A context flow f has an earthquake profile if it has the following properties (earthquake profile): (1) It must contain at least one sequence of context in-*

stances $I_k, I_{k+1},...,I_{k+j}$, such that the distance between the first pair of elements is very different from the distance between the last pair. This property forces a context flow to contain violent and smooth variations in the context transitions. (2) Property two forces a context flow to contain context instances, whose distance to the origin oscillates.

Figure 3 graphically illustrates a context flow with an *EP*. In the figure, the abscissa axis represents the occurrence of context instances over time, whereas the ordinate axis represents the distance of such context instances with respect to the origin ⊗. The sequence of instances $I_k, I_{k+1},...,I_{k+j}$ in figure 3 satisfies the first property, because the distance between $I_k$ and $I_{k+1}$ is much smaller than the distance between $I_{k+j-1}$ and $I_{k+j}$. We can also notice that the distance between all the instances and the origin oscillates.

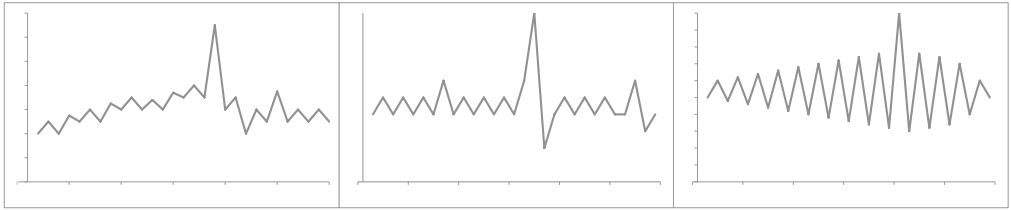

**Figure 4: Examples of the different shapes that an *AEQ* can adopt**

It is worth mentioning that the graphical representation of an earthquake profile can have a variety of shapes, figure 4 illustrate three of them. At the left it shows an *AEQ* with smooth transitions but incremental distance between the instances and the origin. At the center it shows an *AEQ* with smooth transitions with approximately equal distance between the context instances and the origin. Finally, at the right it shows an *AEQ* with transitions whose distance incrementally increase until reaching a pick.

An AEQ with a particular shape may uncover a particular set of faults. For instance, the first one can detect faults due to defective handling of context transitions incrementally far from the origin. That is, subtle and smooth transitions between instances that are not very far from each other, but globally increase their distance to the origin. The second one can detect faults due to defective handling of *oscillating* context transitions, which are continuous transitions between context instances relatively close. The distance between the oscillating instances defines if this profile is capable of detecting faults on smooth context transitions. Finally, the last one can detect faults due to defective handling of incrementally violent context transitions. That is, transitions between context instances increasingly far from each other.

### 6.1 Artificial Earthquake synthesis

ASTT comprises the generation and use of artificial earthquakes to uncover faults. Generating *AEQ* consists in selecting context flows containing at least one occurrence of an *EP*. That is, explore the context flow space *searching* for a context flow with the properties described in definition 5.4. Additionally, the candidate flows must represent a particular set of the environment qualities, i.e. they must fulfill an adequacy criterion.

When searching for AEQs we must also consider that, in the real world, context instances generally do not occur randomly, and thus, the candidate flow should be as similar as possible to the real occurrence of context instances. Typically the occurrence of context instances over time is described by the aid of a probabilistic distribution function, and therefore, the candidate context flow must fit such distribution.

Our strategy to generate *AEQ* consists in translating the selection of context instances and the assembly of context flows into a functional optimization or search problem. In this way, a search algorithm such as *genetic algorithm*, or *ant colony optimization* will search a compromise between the different goals of an AEQ, (1) the presence of *EP*, (2) the similarity with reality, (3) and the satisfaction of a particular criterion. Moreover, it is possible that no single *AEQ* completely satisfies the search goals; hence, the search should select as much *AEQ* as needed to accomplish the goals.

We model this search problem as a two-fold optimization. At the top level, the objective is selecting the smallest set of *AEQ* that maximizes coverage of a particular criterion, and the number of *AEQ* with different shapes. At the bottom level, the objective is selecting the *AEQ* that maximizes the occurrences of earthquake profile, and better approximates the real occurrence of the instances over time.

These two-folds help us defining a global, and a local optimization functions based on the following elements.

DEFINITION 5.5 *A coverage criterion $\mu$ captures a set of context instances that must be covered by a context flow (or a set of them). Let $C_L: F \times \mu \to \mathbb{Z}$ be a function mapping context flows to an integer value indicating the number of elements of $\mu$ covered by a single context flow. Let $C_G: \wp(F) \times \mu \to \mathbb{Z}$ be a function mapping a set of context flows to an integer value indicating the number of elements of $\mu$ covered by set of context flows.*

Examples of coverage criteria are *the coverage of all the pairs of instances, the coverage of all the transitions between the pairs of instance, coverage of X% of the transitions,* etc. Such criteria drive the generation of *AEQ*, forcing the search on the uncovered context space.

DEFINITION 5.6 *Let $EP : F \to \mathbb{R}$ a function mapping context flows to a real value indicating the amount of occurrences of EP (property 1 definition 5.4) in a single context flow.*

DEFINITION 5.7 *Let $S : \wp(F) \to \mathbb{Z}$ a function mapping context flows to an integer value indicating the amount of different EP shapes in a sequence of context flows.*

DEFINITION 5.8 *A probabilistic distribution $\varpi$ defines the real occurrence of context instances over time. Let RE: F × $\varpi$ → $\mathbb{R}$ a function mapping context flows to a real value indicating the distance between the actual context flow and the real occurrence of the instances in the context.*

Globally, the optimization goal consists in finding the maximum value for a global function to optimize.

DEFINITION 5.9 *Given $C_G$, S, and the set of context flows **sf** we define the global function to optimize:*

$$G\ (sf) = w_0 * C_G\ (sf, \mu) + w_1 * S\ (sf) - sizeof\ (sf),\ w_0 + w_1 = 1$$

Where $w_0$ and $w_1$ are the respective weight of covering a target criterion and having different *AEQ* shapes. Greater the value of $w_0$ or $w_1$, greater is the importance of optimizing that particular goal. *G (sf)* aims at optimizing the coverage of a criterion, and *AEQ* shapes of a set of *AEQ* while minimizing the number of *AEQs*.

Locally, the optimization goal consists in finding the maximum value for a local function to optimize.

DEFINITION 5.10 *Given $C_L$, EP, RE, and a context **f** we define the local function to optimize:*

$$L\ (f) = w_0 * C_L\ (f, \mu) + w_1 * EP\ (f) + w_2 * RE\ (f, \varpi),$$
$$w_0 + w_1 + w_2 = 1$$

Where $w_0$, $w_1$, and $w_2$ are the respective weight of the elements of a particular criterion covered by the context flow, the amount of occurrences of *EP* and distance of the approach of the context flow with the reality. This function searches to optimize the trade-off between the different search goals when selecting a particular *AEQ*.

Given the global and local optimization functions we propose an algorithm for searching the optimal sequence of *AEQ* for a given context. Since any local search meta-heuristic such as *tabu-search* or *simulated annealing* can be adapted for performing the local search, we describe only the global search algorithm. In this paper we decided to use tabu-search [14] for local search.

Listing 2 shows the pseudo code of our global search algorithm. Initially, two memory structures are created (lines 2-3). The objective of the first memory structure MEM is storing context flows that could be useful in the future. Each element in MEM has an associated iteration number, which is updated iteratively (line 21). The second memory structure T stores the shapes and criterion elements already covered by the candidate solution. It is used by the local search to avoid exploring the areas already covered by existing solution.

  **1:** *procedure globalSearch :* **SOL***: Sequence[ContextFlow]*
  **2:**  **MEM***: Set[ContextFlow]*
  **3:**  **T***: MemoryStructure*

```
 4:   while stop criterion not met do
 5:     for each Æ in MEM do
 6:       if G(SOL U Æ ) > G(SOL) then
 7:         add Æ to SOL
 8:       else
 9:         if maximal amount of iterations of Æ then
11:           remove Æ from MEM
12:         end if
13:     end for each
14:     Æ <- localSearch(T)
15:     if G(SOL U Æ ) > G(SOL) then
16:       add Æ to SOL
17:     else
18:       add Æ to MEM
19:     end if
20:     update T with the new elements in SOL
21:     update iteration number of each element in MEM
22:   end while
23: end procedure
```

**Listing 2: Global search algorithm for the generation of *AEQ***

Once the data structures are initialized and while a stop criterion such as a minimum value for *G* or a maximal number of iterations is not met, the algorithm proceeds (line 4). Otherwise, it returns the candidate solution SOL. For each AEQ in MEM, the algorithm evaluates the utility of adding it to the candidate solution (line 6). If adding it increases the value of *G*, then it is added to the candidate solution (line 7). Otherwise, if it has reached a maximum amount of iterations in the memory structure MEM it is deleted (line 11). Next, a local search algorithm generates an *AEQ* that does not overlap the elements in T. The algorithm evaluates the utility of adding it to the candidate solution. If adding it increases the value of *G*, then it is added to the candidate solution (line 7). Otherwise, it is added to the memory structure MEM with an initial iteration count of 0. Finally, the data structure T is updated with the new elements in the candidate solution, and the MEM iteration number is updated.

We have implemented this algorithm as well as a taboo local search as a ~7000 LOC java program we refer as *shaker*.

## 7 Experiments

In the previous section we presented *ASTT* and proposed a strategy to synthesize *AEQ* (context flows with particular properties). Our hypothesis is that such data may help testers finding faults in the adaptation policy specification and its realization. This section describes the empirical evaluation of this hypothesis. Section 6.1 presents our test subject and describes the instrumentation we per-formed in order to obtain the experimental data. Section 6.2 describes the setting of the experiment. Finally section 6.3 presents and discusses the results.

## 7.1 Test subject

In order to validate our hypothesis about the ability of *ASTT* to uncover faults in adaptation policies, we use the adaptive web server presented in section 2 as a test subject.

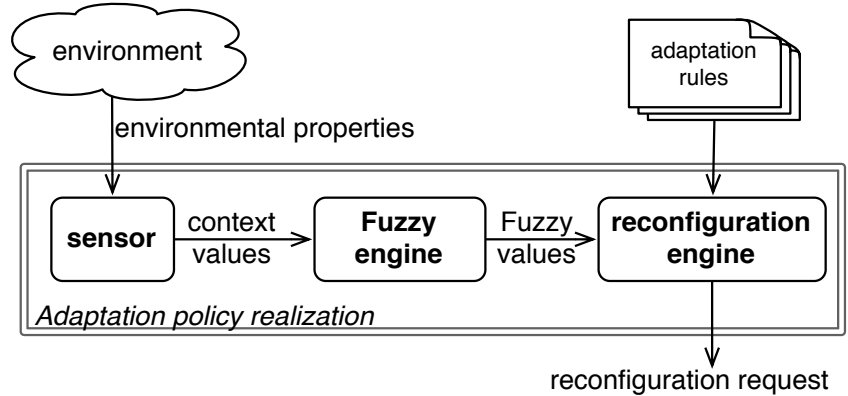

**Figure 5: Architecture of the adaptive web server adaptation policy realization**

Figure 5 presents the architectural realization of the adaptation policy presented in section 2. It is composed of a *sensor component*, which is aware about the environment and collects the data produced by environmental changes. It encodes the data into values representing the environmental properties of interest (context instance) and passes them to a *fuzzy engine*. The *fuzzy engine* converts these values into fuzzy values (adjectives such as *high*, *low*, or *medium*) and passes them to a *reconfiguration engine*. Finally, the *reconfiguration engine* loads the adaptation rules and matches the fuzzy values against the adaptation rules. If an adaptation rule matches the values, then it requests the system implementation to reconfigure as described by the rule.

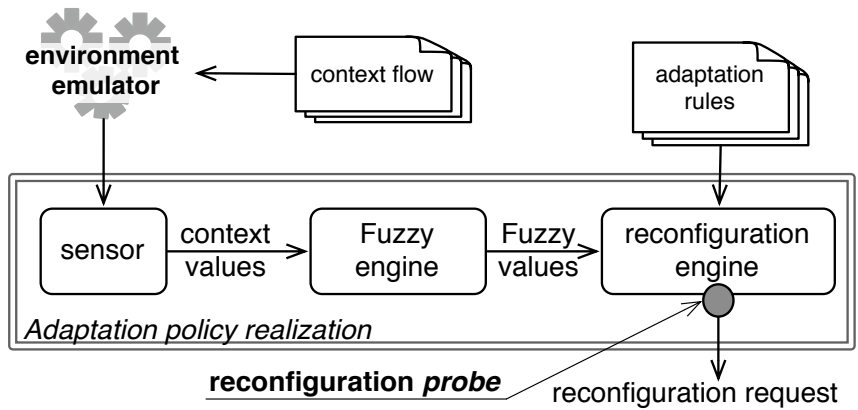

**Figure 6: Instrumented architecture of the adaptive web server adaptation policy realization**

In order to inject context instances and collect reconfiguration data we have instrumented the adaptation policy realization. Figure 6 presents the instrumented architecture. We have modified the source code of the *sensor component* and replaced the environment sensing mechanism with an *environment emulator*. This emulator reads context flows from a text file and injects them into the system provoking the instrumented sensor to respond identically to the non-instrumented one. We have also added a *reconfiguration probe* that records the reconfiguration requests produced by the reconfiguration engine. These requests constitute a variant flow as described in section 2.

### 7.2 Experiment set up

We prepared and executed our experiment in the following way. (1) Initially we introduced a series of faults into the adaptation policy realization generating a set of *mutant* versions of it. (2) Next using the algorithm described in section 5.1 we generated 3 series of *AEQ*s (test data). (3) Finally, we executed the system exposing it to the environmental variations described by the *AEQ*s. We collected and compared the traces generated by the reconfiguration probes in the original and mutant versions of the web server. Whenever the traces produced by the original version where different from the mutant version we declared that the test data killed that mutant. In the reminder of this section we detail the faults we introduced into the instrumented adaptation policy, the settings we used to generate the sets of *AEQ*s, and finally we detail the support we used to execute the mutants and evaluate their results.

### 7.3 Mutants

We have introduced 90 faults into instrumented version of the adaptation policy realization generating 90 *mutants* of it. We classify the introduced faults in 4 groups:

*F1.* Faults introduced in the values transmitted from the sensor component to the fuzzy engine (3 faults). This fault consists in changing the order and magnitude of the property values of each context instance.

*F2.* Faults introduced in the calculation of the fuzzy values in the fuzzy engine (25 faults). These faults consist in permuting the fuzzy values passed from the fuzzy engine to the reconfigurations engine. For instance, whether the values (adjectives) passed from the fuzzy engine were *high, low,* and *medium*, we replaced *high* by *low* and *low* by *high*.

*F3.* Faults introduced in the adaptation policies leaving gaps in the adaptation (34 faults). These faults consist in changing the adaptation rules fuzzy values, one each time. For instance, on listing 1 in section 2.2 we changed the value *high* by *low* (line 5). Such change leaves a gap in the possible events the adaptation rules capture; in this case when the fuzzy value is *high* no action is performed.

*F4.* Faults introduced in the adaptation rules without leaving gaps in the adaptation (28 faults). These faults are similar to *F3*, however, instead of changing one value each time, we permuted pairs of fuzzy values. For instance, on listing 1 in section 2.2 we changed the value *high* by *low* on the first rule (line 1), and *low* by *high* on the second (line 5). These changes invert the action that may take place when a fuzzy value occurs for a given property.

### 7.4 Data generation parameter settings

Using *shaker*, our java implementation of the algorithm described in section 5.3, we have generated 3 sets of 100 *AEQ*. For this experiment we decided to fix the length of each *AEQ* to 60 context instances.

Concerning the coverage of the generated data, we used pairwise testing techniques [4, 22] to generate *all the possible context instance pairs*, and employ them as a coverage indicator. That is, try to cover the entire valid context instances generated by the possible pair combination of their property values. We calculated such combinations using the *AllPairs*[5] tool. Since this tool does not take into account the context constraints when generating the pairs, we filtered them extracting those that satisfied the constraints. For example, we removed the context instance <5,100,1> because having more files than requests is not relevant.

On the global search, we fixed the permanence of each *AEQ* in the MEM structure to 10 iterations. We implemented the memory structure *T* (section 5.1, listing 2, line 3) as a list containing the instances already covered by the candidate solution. We established as stop criterion a total of 100 iterations with no amelioration of the *G* value. Besides, whether the *G* value continued increasing, we fixed a hard limit of 1000 iterations.

Regarding the local search, we used a tabu-search meta-heuristic with a simple tabu list [14] of size 30 (half of the *AEQ* length). The movement we implemented for this search consisted in the changing the property values of a single context instance in order to increase or decrease the distance with its neighbors. In order to minimize the amount of overlap in the criterion coverage, we matched the elements memorized by the global search against the elements generated by the tabu-search. Whenever a maximum of three context instances were in the memory, we forced the algorithm to move the solution away from them.

We have parameterized the global and local optimization functions in the following way. In the global optimization function *G*, we have assigned the same importance to the coverage criterion and the *EP* shape (definition 5.9, $w_0=0.5$, $w_1=0.5$). In the local optimization function, we have assigned a major importance to the occurrence of an *EP* (definition 5.10, $w_0=0.6$), and a relatively minor importance to the coverage criterion (definition 5.10, $w_1=0.4$). Notice that we have disregarded the similitude of the *AEQ* with the reality (definition 5.10, $w_3=0$); the rationale for doing so is that file requests can arrive with a very large range of probabilistic distributions over time. Such distribution depends on the application domain of the web server such as online sales, content management, etc. Since the adaptive web server is intended to work on multiple domains, we decided to disregard it.

---

[5] http://www.mcdowella.demon.co.uk/allPairs.html

### 7.5 Execution

We have simulated the environmental variations drawn by the 3 sets of *AEQs* over the 90 mutants. To do so we executed the initially instrumented policy realization, as well as the 90 mutants on a grid composed of 90 computers equipped each with two *Intel™ Xeon* processors at 3.4Ghz, and 4 Gb of main memory. The execution of the 27300 simulations (91 program * 100 simulation/AEQ * 3 AEQ/program) took about 55 minutes, and each computer executed in average 600 simulations.

Once the simulations were completed, we compared the traces (variant flows) produced by each mutant with those produced by the initial policy realization. We performed such comparison using a custom program written in java, which interprets and compares each trace with the system configuration at each point where a context instance arrives. This tool enabled us to determine the precise points where the adaptation policy produced the wrong configuration (with respect to the initial realization). We then say that whenever the simulation of an *AEQ* on a mutant $t$ produces a trace that differs from the trace produced by the original server, it kills the mutant $t$.

In the next section we present and discuss the results of comparing of the traces produced by the 27000 simulations.

### 7.6 Results and analysis

Figure 7 presents a chart containing the average amount of the 3 sets of *AEQ* killing each mutant. Vertical bars represent the different mutants from 1 to 90, their color indicated the different groups of faults they belong to, and the value at left is the average amount of *AEQ* killing the mutants. This chart helps us reasoning about to illustrate whether *ASTT* was capable of detecting faults, and if some of these faults were more or less difficult to detect.

In general the sets of *AEQ* were capable of killing the 96% of the mutants. Nevertheless, we analyzed the survival mutants noticing that they were equivalent with the original adaptation policy realization. This means that our test data was capable of killing the 100% of the mutants. This result is positive because it indicates that *AEQ*s were capable of finding each fault we introduced. Notice that 56% of the mutants were killed by all the *AEQs*, and that 4% by more than 60% of the *AEQs*.

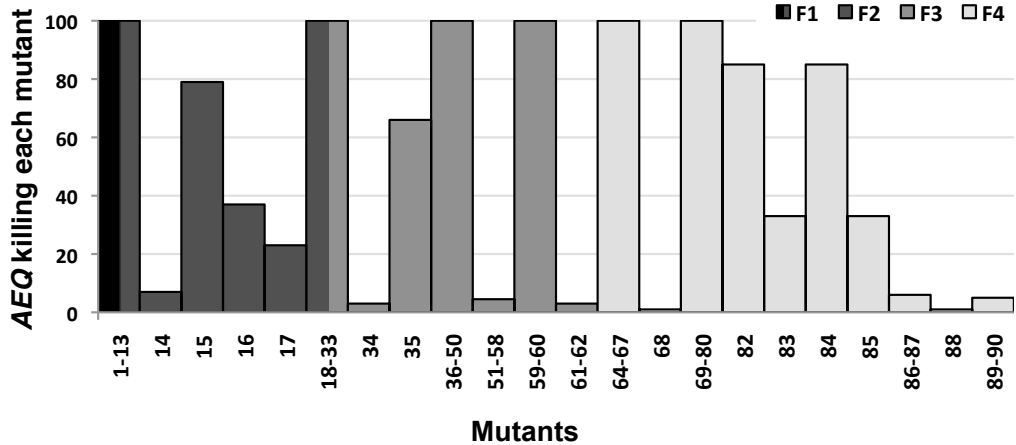

**Figure 7: Amount of *AEQ* killing each mutant**

In the following we analyze the results by each group of faults. At this point it is worth mentioning that the adaptive web server adaptation policy encodes property values using adjectives such as *low, high, and medium.*

*F1:* The totality of *AEQ*s was capable of killing the mutants of this group. The rationale behind this success is that mutants realizing these faults are very sensitive to environmental variations. Since this group consists in permuting environmental properties with different domains (*request density, request dispersion*), the variations introduced by *AEQ*s always produced wrong configurations. Therefore, this group of faults does not help us determining whether *AEQ* can actually find more subtle faults in the adaptation policy.

*F2:* The mutants of this group are those introducing permutations between low and high for the *request density*, and *request dispersion* properties. They are killed by more than 60% of the generated *AEQ*s. Interestingly, the mutant 14 permuted medium by low values of the *request dispersion* property. This fault can be detected only by AEQs that contain smooth changes on the property *request dispersion*. Mutants 16 and 17 show another interesting result. They introduced permutations were high values of the property *request density* are always replacing medium values. This fault can only be detected by smooth variations on the *request density* property. The low amount of *AEQ*s killing the last mutants indicates that only a few of them contain smooth changes between high and low values of the properties *request density* and *request dispersion*.

*F3:* The mutants of this group, killed by more than 60% *AEQ*s were those introducing gaps on the adaptation to violent environmental changes. The major part of such gaps consists in replacing the adaptation facing high values by those facing low values (and vice versa) of the *request density* property. Consequence of this, *AEQ*s introducing violent context variations from low to high values on this property were able to detect these faults. A different situation occurs with faults replacing medium by high, and low values on the *request dispersion* and *request density* properties (Mutants 34, 51-58). This fault can only be detected by smooth environmental variations passing from high to medium, and low to medium values. Mutants 61, 62 produce another interesting result. They replace the high values of the *request density* property by medium or low values. These values were used by the adaptation policy handling the removal of data servers. This fault is particularly dependant of the system history, and is sensible only to context changes introducing initially low values, followed by a high and a low value.

*F4:* Analogous to the previous group, in this group the mutants killed by more than 60% *AEQ*s were those permuting low to high, and high to low values in the *request density* and *request dispersion* properties. We pay special attention to mutant 68, which permutes high by medium values in the adaptation handling the removal of cache. This fault is sensible only to context changes with initially high values, followed by low values on both, *request density* and *request dispersion* properties. Mutants 83 and 85 are a particular case of the fault introduced by mutant 16 and 17. They permute the values high and low by middle on the *request density* property. More precisely, the property values used by the adaptation handling the deployment of data servers. The faults introduced by mutants 86, 87 are equivalent to those introduced by mutants 89, 90. These faults permute medium by high and low values in the rule stating the removal of data servers. Only smooth environmental variations from high to medium request density can detect such faults.

The results obtained on each group of faults allows us to infer the following conclusions:

1. The experimental data obtained for fault groups *F2, F3*, and *F4* supports our hypothesis: *ASTT can detect faults in DAS's adaptation policy*. Evidence of this is the high amount of mutants killed, as well as the high percentage of *AEQ*s killing mutants that introduce faults affecting the handle of violent environmental variations. Furthermore, the experiments show evidence that several *AEQ*s are needed in order to detect different types of faults.
2. Although *AEQ*s were initially meant to detect faults caused by wrong handling of violent context variations, and not particularly smooth variations, experimental evidence show that they were capable of detecting such faults. This is explained by the fact that the *AEQ* generation algorithm allows the generation *AEQ*s containing smooth context changes. That is, *AEQ*s with different shapes.
3. A number of the introduced faults were sensible only to particular sequences of context instances. The empirical evidence demonstrates that *AEQ*s contained such sequences, and that the order in which the instances composing an AEQ must be assembled cannot be disregarded.
4. The parameter setting of the AEQ generation algorithm on listing 2 produced *AEQ*s with tendency to particularly violent context variations. Evidence of this is the large portion of test cases detecting faults caused by wrong handling of violent context changes. Furthermore, we were capable of automatically generating particular *AEQs* with different EP shapes, and that globally cover all the context pairs (pairwise testing criterion).

### 7.7 Threats to validity

There are three threats to the validity of our experiments. The first comes from the application of our strategy to only a single test subject. In order to make more general statements about the effectiveness of ASTT, it will be necessary to apply the strategy and algorithms introduced in this paper to a large scope of *DAS*. We plan to do so in the context of the European project DiVa [1], which comprises two large case studies.

The second threat comes from the use of only one parameter setting for the experiment. We have generated 3 sets of AEQ using the identical parameter settings for the generation algorithm. This implies that the 3 test sets produce approximately the same results, which allows us to make statements about the average results. However, in order to make more precise statements about the effectiveness of *ASTT,* it will be necessary to generate *AEQ*s with a variety of parameters. Particular threats to validity are the length of each *AEQ* and the parameterization of global, and local optimization functions. The length of each *AEQ* can affect the number of *AEQ* needed to find a particular fault, such as those requiring smooth context variation, and the time consumption of the tests. Moreover, the compromise between the coverage of a particular criterion and the *EAQ* shape can also affect the number of *AEQ*s that can detect a particular fault.

Finally, the third threat to validity comes from the choice of the faults we introduced. We did not introduce every possible fault into the adaptive web server. Instead we introduced the faults we thought meaningful to our case study, such as modification in the adaptation policy bearing violent and smooth context variations. Considering every possible mutant will allow testers to precisely identify the faults that *ASTT* is more, or less suitable to find.

## 8 Related work

A number of researchers have addressed the validation of adaptive systems. Zhang et al. [25] address the verification of dynamically adaptive systems through modular model checking. For each transition between systems variants, they model check only the parts of the system that have change product of an adaptation. In [24], they introduce a model-based adaptive software development process that uses Petri nets to model the behavior, and uses existing Petri net-based model checking tools to verify these models gain interesting properties. Kramer and Magee [19] use property automata to specify the properties of adaptive program, and labeled transition system analysis to verify these properties. These works diverges form ours because instead of verifying the system and its adaptations, we propose to validate the adaptation driver (adaptation policy) independently from the underlying platform. Besides they are founded on formal methods and verification, whereas our work on testing techniques.

Lu et al. [16] study the test of pervasive context-aware software. They assume context awareness as a series of *if-then* cases, and starting from that point they formalize the notions of context aware data flow entities, i.e. entities that manipulate data coming from the context. By using this formalization they propose a family of test adequacy criteria that measure the quality of test sets with respect to the context variability. The underlying idea of this work is pretty similar to ours *testing the driver of potential adaptations*. However we do not perform any data flow analysis on the context data, and our proposition can address a much larger variety of reasoning strategies, including those relying on the system state.

Combinatorial interaction testing [4] consists in sampling a test data space in such a way that its *t*-possible combinations are included; pairwise or 2-way combinations are the most commonly studied. Many researchers [6-8, 15, 22] have explored the generation of such combinations with the prime objective of producing the smallest subset of test data to achieve the desired t-way. Although combinatorial testing is efficient reducing the size of test to run, it is not sufficient for testing adaptation policies. This is because besides considering the context space, it is necessary to consider the transitions between the elements of such space (flow space). The benefit of *AEQ*s over selecting the *t*-wise is that they ensure the presence of specific properties that targeting specific kind of fault. Furthermore, even if the *t*-wise is not enough for testing the adaptation policy, we have used the pairwise as a coverage criterion.

Search based testing consists in the use of random or directed search techniques (hill climbing, genetic algorithms etc.) to address problems in the software testing domain [20]. Our contribution comes to form part of such body of work because we search to generate the better set of *AEQ*s capable of finding faults in adaptation policies.

The combinatorial test of software product lines also relates our work. Since it searches to check whether a set of product is valid we think it can be used in combination to our strategy. Cohen et al. [5] study the coverage and adequacy criteria for testing software product lines. They propose mapping a variability model into a simple relational model that satisfies the requirements of interaction testing. In this way the relational model is used as a covering array, which defines test adequacy and coverage criteria.

## 9 Conclusions and perspectives

Testing whether adaptation policies are correctly implemented and well suited for their working environment is challenging not because of the process itself, but for the large amount of testing data.

Although simulating environmental changes is not particularly hard, it is hard to simulate each possible environmental condition and the transitions between them (environmental space). That is because the amount of possible environmental conditions grows exponentially with each environmental property. This drawn the simulation of all the environmental conditions and their transition not feasible in a reasonable amount of time.

In this paper we proposed a strategy for selecting only a portion of the environmental space. Inspired by a civil engineering technique called *shaking table testing*, we proposed *artificial shaking table testing* (*ASTT*). *ASTT* put forward the use of *artificial earthquakes* (*AEQ*) to test the *resistance* of adaptation policies to violent environmental changes. Basically, an *AEQ* is a sequences of environmental conditions characterized by an *earthquake profile (*EP*)*, which is the presence of violent variation in the transitions between each condition in the sequence. Our hypothesis was that *ASTT* is capable of detecting faults due in the adaptation policy implementation and design. More precisely, faults that are due to erroneous specification or handling of violent environmental changes.

We automated the generation of *AEQ* by translating their formulation into a search problem and defining two optimization functions. Next, we proposed an initial algorithm to explore the environmental space. The experimental results exhibit evidence that corroborate our hypothesis. Out of 90 faults introduced into an adaptation policy realization, *ASTT* was capable of detecting the 100% of them. *AEQ*s resulted to be particularly good in the detection of faults lying on violent environmental changes. Furthermore, the experiments show that *AEQ*s are also capable of detecting faults due to smooth environmental variations.

The benefits of using *ASTT* for testing adaptation policies are various. It can help testers uncovering faults related to violent and smooth environmental changes. Furthermore, it can help testers uncover design faults in adaptation policies specification and assess their adequacy with respect to their working environment.

In future work we plan experimenting with different case studies, particularly large scale dynamically adaptive systems. This will give use a more precise indication of the scalability and effectiveness of *ASTT*. We also plan improving the algorithms for local and global search, and study the use of different coverage criteria in the generation of *AEQ*s.

## 10 ACKNOWLEDGMENTS

This work was partially supported by the European project DiVA (EU FP7 STREP).